\documentclass[twocolumn,pra,showpacs,superscriptaddress,amssymb,amsmath,amsmath]{revtex4-1}
\usepackage{graphicx}
\usepackage{epstopdf}
\usepackage{bm}
\usepackage{hyperref}
\usepackage{comment}
\usepackage[version=3]{mhchem}
\usepackage{multirow}
\usepackage{float}

\hypersetup{%
   pdfpagemode=None, 
   pdfstartpage=1,
   pdfmenubar=true,
   pdftoolbar=true,
   colorlinks = true,
   linkcolor=blue,
   citecolor=blue,
   urlcolor=blue,
   bookmarksopen=false
 }

\newcommand{\be}{\begin{equation}}
\newcommand{\ee}{\end{equation}}

\begin{document}
\title{Magnetically tunable Feshbach resonances in ultracold gases of europium atoms \\ and mixtures of europium and alkali-metal atoms}

\author{Klaudia Zaremba-Kopczyk}
\affiliation{Faculty of Physics, University of Warsaw, Pasteura 5, 02-093 Warsaw, Poland}
\author{Piotr S. \.Zuchowski}
\affiliation{Faculty of Physics, Astronomy and Informatics, Nicolaus Copernicus University in Torun, Grudziadzka 5, 87-100 Toru\'n, Poland}
\author{Micha\l~Tomza}
\email{michal.tomza@fuw.edu.pl}
\affiliation{Faculty of Physics, University of Warsaw, Pasteura 5, 02-093 Warsaw, Poland}
\date{\today}

\begin{abstract}

We investigate magnetically tunable Feshbach resonances between ultracold europium atoms and between europium and alkali-metal atoms using multichannel quantum scattering calculations. For ultracold gases of europium atoms both homonuclear $^{153}$Eu+$^{153}$Eu and heteronuclear $^{151}$Eu+$^{153}$Eu systems are studied. Calculations for mixtures of europium and alkali-metal atoms are carried out for prototype systems of $^{153}$Eu+$^{87}$Rb and $^{153}$Eu+$^7$Li. We analyze the prospects for the control of scattering properties, observation of quantum chaotic behavior, and magnetoassociation into ultracold polar and paramagnetic molecules. We show that favorable resonances can be expected at experimentally feasible magnetic-field strengths below 1000$\,$G for all investigated atomic combinations. For Eu atoms, a rich spectrum of resonances is expected as a result of the competition between relatively weak short-range spin-exchange and strong long-range magnetic dipole-dipole interactions, where the dipolar interaction induces measurable resonances. A high density of resonances is expected at magnetic-field strengths below 200$\,$G without pronounced quantum chaos signatures. The present results may be useful for the realization and application of dipolar atomic and molecular quantum gases based on europium atoms in many-body physics.

\end{abstract}

\maketitle

\section{Introduction}

Magnetically tunable Feshbach resonances are a universal and useful tool to control collisional properties in ultracold quantum gases~\cite{JulienneRMP06a,JulienneRMP10}. They have been essential for the realization of a plethora of ground-breaking experiments in quantum many-body physics~\cite{BlochRMP08,BlochNP12}. Magnetic Feshbach resonances are expected between any open-shell atoms, but first applications involved ultracold alkali-metal atoms~\cite{InouyeNature98}. Nevertheless, they were also observed and employed in experiments with ultracold Cr atoms~\cite{WernerPRL05,PfauPRL05,PavlovicPRA05,LahayeNature07,KochNP08}, and recently with ultracold Er and Dy atoms~\cite{LevPRL11,LevPRL12,FerlainoPRL12,AikawaPRL14,PetrovPRL12,MaierPRX15,MaierPRA15}. Moreover, they were measured in mixtures of Yb atoms in the metastable $^3P$ state with the ground state Yb~\cite{KatoPRL13,HoferPRL15} or Li~\cite{KhramovPRL14} atoms, and in a mixture of the ground-state closed-shell Sr and open-shell Rb atoms~\cite{Barbe2017,ZuchowskiPRL10}. 

Ultracold gases of dipolar atoms are especially interesting because the rich physics of different quantum phases and spin models can be realized with them~\cite{LahayeRPP09,BaranovCR12}. Therefore, atoms in complex electronic states with large both spin and orbital electronic angular momenta, such as Er and Dy, have been cooled down to low and ultralow temperatures. Tremendous successes have already been accomplished with these atoms, just to mention the observation of quantum chaos in ultracold collisions~\cite{FrischNature14}, Fermi surface deformation~\cite{AikawaScience14}, self-bound quantum droplets~\cite{SchmittNature16}, Rosensweig instability~\cite{KadauNature16}, and extended Bose-Hubbard models~\cite{BaierScience16}. The spin dynamics of impurities in a bath of strongly magnetic atoms and magnetic polaron physics~\cite{AshidaPRB18,Ardila2018} wait for realization.

The first highly dipolar atoms obtained at ultralow temperatures were Cr$\,$($^7S_3$), Dy$\,$($^5I_8$), and Er$\,$($^3H_6$), however, several other transition-metal or lanthanide atoms may potentially be used. For example, magneto-optical cooling and trapping of Tm$\,$($^2F_{7/2}$)~\cite{SukachevPRA10} and Ho$\,$($^4I_{15/2}$)~\cite{MiaoPRA14} were also realized. Another lanthanide candidate is Eu$\,$($^8S_{7/2}$). The buffer-gas cooling and magnetic trapping of Eu atoms were demonstrated~\cite{DoylePRL97,DoyleNature04,SuleimanovPRA10}, and recently magneto-optical cooling and trapping of optically pumped metastable Eu$\,({}^{10}D_{13/2})$ atoms were achieved~\cite{Inoue2018}. Further cooling to the quantum degeneracy should not be more challenging than the already demonstrated production of ultracold gases of other lanthanide atoms with more complex electronic structure~\cite{LevPRL11,LevPRL12,FerlainoPRL12,AikawaPRL14}. In contrast to Er and Dy atoms, ground-state Eu atoms do not have any electronic orbital angular momentum ($l=0$), and their large magnetic dipole moment is solely related to the large electronic spin angular momentum ($s=7/2$) of seven unpaired $f$-shell electrons. Eu atoms, thus, are more similar to Cr atoms than to other lanthanides. However, they possess 17\% larger dipole moment than Cr, which combined with three times larger mass of Eu as compared to Cr will result in four times stronger dipole-dipole interactions in ultracold gases of Eu atoms as compared to Cr atoms, but four times weaker interactions as compared to Dy and Er atoms (the strength of the dipolar interaction is $a_\text{dd}\sim d^2m$, where $d$ is the dipole moment and $m$ is the mass of atoms~\cite{LahayeRPP09}).

Heteronuclear molecules possessing a permanent electric dipole moment are another promising candidate for numerous applications, ranging from ultracold controlled chemistry to quantum computation and quantum simulation of many-body physics~\cite{CarrNJP09,QuemenerCR12,BohnScience17}. Heteronuclear molecules formed of atoms with large magnetic dipole moments could possess large both magnetic and electric dipole moments useful for investigating the interplay between the electric and magnetic dipolar interactions and phases in ultracold gases. Therefore, the chromium--alkali-metal-atom molecules such as CrRb~\cite{SadeghpourPRA10}, chromium--closed-shell-atom molecules such as CrSr and CrYb~\cite{TomzaPRA13a}, europium--alkali-metal-atom molecules such as EuK, EuRb, and EuCs~\cite{TomzaPRA14}, and erbium-lithium molecules ErLi~\cite{GonzalezPRA15} were theoretically investigated and shown to possess large both electric and magnetic dipole moments. Experimentally, the magnetoassociation into ultracold Er$_2$ dimers~\cite{FrischPRL15} and photoassociation into spin-polarized Cr$_2$ dimers~\cite{RuhrigPRA16} were demonstrated. Ultracold mixtures of Dy and K atoms~\cite{Ravensbergen2018} and Dy and Er atoms~\cite{IlzhoferPRA18} were also obtained, opening the way for the formation of ultracold highly magnetic and polar molecules in nontrivial electronic states. The extraordinarily rich, dense, and chaotic spectra of magnetic Feshbach resonances for Dy and Er atoms~\cite{KotochigovaRPP14,MaierPRX15,MaierPRA15} may, however, make the magnetoassociation into heteronuclear molecules and investigation of magnetic polaron or Efimov physics difficult. The use of ultracold Eu atoms may be a remedy.  

Here we investigate magnetically tunable Feshbach resonances between ultracold europium atoms and between europium and alkali-metal atoms using multichannel quantum scattering calculations. We study both homonuclear $^{153}$Eu+$^{153}$Eu and heteronuclear $^{151}$Eu+$^{153}$Eu systems of europium atoms, and $^{153}$Eu+$^{87}$Rb and $^{153}$Eu+$^7$Li combinations as prototype systems of mixtures of europium and alkali-metal atoms. We show that resonances favorable for the control of scattering properties and magnetoassociation into ultracold polar and paramagnetic molecules can be expected at experimentally feasible magnetic field strengths below 1000$\,$G for all investigated atomic combinations. The density of $s$-wave resonances strongly depends on the projection of the total angular momentum on the magnetic field. For Eu atoms, the dipolar interaction induces measurable resonances, and a high density of resonances without pronounced quantum chaos signatures is expected at magnetic field strengths below 200$\,$G.

The plan of this paper is as follows. Section~\ref{sec:theory} describes the used theoretical methods. Section~\ref{sec:results} presents and discusses the numerical results and physical implications of our findings. Section~\ref{sec:summary} summarizes our paper and presents future possible applications and extensions.

\section{Computational details}
\label{sec:theory}

\begin{figure}[tb!]
\begin{center}
\includegraphics[width=1\linewidth]{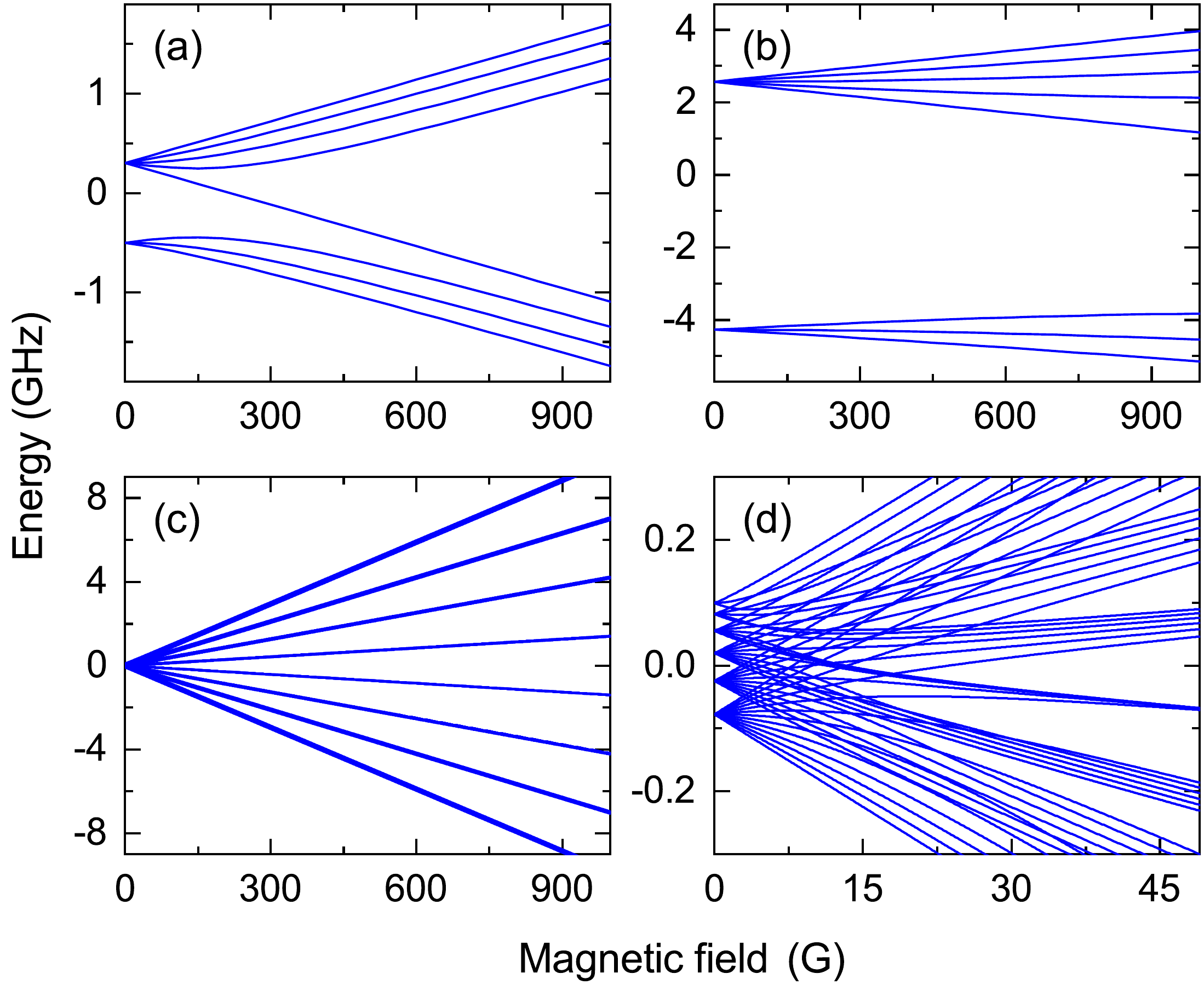}
\end{center}
\caption{Hyperfine energy levels for (a)~\ce{^7Li}, (b)~\ce{^87Rb}, and (c,d)~\ce{^153Eu} atoms as a function of the magnetic field. Panel~(d) shows an enlarged part of panel~(c) for the small magnetic field.}
\label{fig:hyperfine}
\end{figure}

Europium atoms in the electronic ground state have very large electronic spin angular momentum ($s=7/2$), but they do not have any electronic orbital angular momentum ($l=0$). This results in the $^8S_{7/2}$ term. Lithium and rubidium atoms, as all alkali-metal atoms, have simpler structures described by the $^2S_{1/2}$ term. Characteristics of all investigated atoms are collected in Table~\ref{tab:atoms}, and atomic hyperfine energy levels as a function of the magnetic field are presented for $^7$Li, $^{87}$Rb, and $^{153}$Eu in Fig.~\ref{fig:hyperfine}. Coupling of the electronic spin with the nuclear spin, which is $i=5/2$ for both isotopes of Eu, results in a very rich hyperfine structure for these atoms. Interestingly, hyperfine coupling constants for Eu are small and negative. They are between one to two orders of magnitude smaller than for alkali-metal atoms; therefore, the regime dominated by the Zeeman interaction with the linear dependence of hyperfine energy levels on the magnetic field can be observed for Eu atoms at relatively small strengths of the magnetic field (cf.~Fig.~\ref{fig:hyperfine}). A negative value of the hyperfine coupling constant means the inverse order of hyperfine levels; thus the hyperfine ground state of Eu atoms has angular momentum of $f=6$.

\begin{table}[b!]
\caption{Terms $^{2s+1}l_j$, electronic spins $s$, nuclear spins $i$, possible total angular momenta $f$, and hyperfine coupling constants $a_\text{hf}$ for the investigated atoms. \label{tab:atoms}}
\begin{ruledtabular}
\begin{tabular}{c c c c c c}
	atom &  $^{2s+1}l_j$, & \textit{s} & \textit{i} & \textit{f} & $a_\mathrm{hf}\,$(MHz)\\ 
	\hline
	\ce{^7Li}& $^2S_{1/2}$ & 1/2 & 3/2 & 1, 2 & 401.752~\cite{ArimondoRMP77} \\ 
	\ce{^87Rb}& $^2S_{1/2}$ & 1/2 & 3/2 & 1, 2 & 3417.34~\cite{ArimondoRMP77} \\ 
	\ce{^151Eu}& $^8S_{7/2}$ & 7/2 & 5/2 & 1, ..., 6 & -20.052~\cite{SandarsPRSA60} \\  
	\ce{^153Eu}& $^8S_{7/2}$ & 7/2 & 5/2 & 1, ..., 6  & -8.853~\cite{SandarsPRSA60} \\ 
\end{tabular} 
\end{ruledtabular}
\end{table}

\begin{table*}[tb!]
	\caption{Parameter values of the used Morse/Long-range potential-energy functions fitted to \textit{ab initio} data from Refs.~\cite{TomzaPRA14,BuchachenkoJCP09}. $D_e$ is in cm$^{-1}$ and other parameters are in atomic units or are dimensionless.~\label{tab:mlr}}
	\begin{ruledtabular}
		\begin{tabular}{c c c c c c}
		\multirow{2}{*}{parameter}& \multicolumn{2}{c}{Eu+Li} & \multicolumn{2}{c}{Eu+Rb} & Eu+Eu  \\ 
		 & $V_{S=3}$ & $V_{S=4}$ & $V_{S=3}$ & $V_{S=4}$ & $V_{S=7}$ \\ 
		\hline 
		$D_e$& 2971.0 & 2443.2 & 1239.1 & 1047.1 & 704.32 \\ 
		$R_e$& 6.5561 & 6.7288 & 8.6393 & 8.7904 & 9.2919 \\ 
		$\varphi_0$& -1.1665 & -0.96262 & -0.80406 & -0.75536 & -0.78004 \\ 
		$\varphi_1$& 0.24492 & 0.37826 & -0.28015 & -0.23809 & -0.49552 \\  
		$\varphi_2$& 1.2024 & 0.49243 & -0.71864 & -0.66458 & -0.28324 \\ 
		$\varphi_3$& -1.1880 & -1.3791 & -0.88769 & -0.54075 & 0.36690 \\ 
		$\varphi_4$& -4.2923 & -2.9548 & -1.2472 & -0.84187 & -0.45423 \\ 
		$C_6$ & 2066 & 2066 & 3779 & 3779 & 3610 \\ 
	\end{tabular} 
\end{ruledtabular}
\end{table*}

The Hamiltonian describing the nuclear motion of two colliding atoms, $A+B$, reads
\begin{equation}\label{eq:Ham}
\begin{split}
    \hat{H}=&-\frac{\hbar^2}{2\mu}\frac{1}{R}\frac{d^2}{dR^2}R+
    \frac{\hat{L}^2}{2\mu R^2}+
    \sum_{S,M_S}V_S(R)|S,M_S\rangle\langle S,M_S|\\
    &+\hat{H}_{A}+\hat{H}_{B}+\hat{H}_{ss}\,,
\end{split}
\end{equation}
where $R$ is the interatomic distance, $\hat{L}$ is the rotational angular momentum operator, $\mu$ is the reduced mass, $V_S(R)$ is the potential-energy curve for the state with the total electronic spin $S$, and $|S,M_S\rangle\langle S,M_S|$ is the projection operator on the states with the total electronic spin $S$ and its projection $M_S$. The atomic Hamiltonians, $\hat{H}_j$ ($j=A,B$), including hyperfine and Zeeman interactions are given by
\begin{equation}\label{eq:Ham_at}
\hat{H}_j=a_{j}\hat{i}_{j}\cdot\hat{s}_{j}
  +\left(g_e\mu_{{\rm B}}\hat{s}_{j,z}+g_{j}\mu_{{\rm N}}\hat{i}_{j,z}\right)B_z\,,
\end{equation}
where $\hat{s}_{j}$ and $\hat{i}_{j}$ are the electronic and nuclear spin angular momentum operators, $a_{j}$ is the hyperfine coupling constant, $B_z$ is the magnetic field strength, $g_e$ and $g_j$ are the electron and nuclear $g$ factor, and $\mu_{\text{B}}$ and $\mu_{\text{N}}$ are the Bohr and nuclear magneton, respectively. For Eu atoms we neglect the hyperfine electric quadrupole coupling because it is much smaller than the leading hyperfine magnetic dipole coupling. The magnetic dipole-dipole interaction between electronic spins is
\begin{equation}
\hat{H}_{ss}=\frac{\alpha^2}{R^3}\left(\hat{s}_A\cdot\hat{s}_B - 3 \hat{s}_{A,z}\hat{s}_{B,z} \right)\,,
\end{equation}
where $\alpha$ is the hyperfine coupling constant. 

\begin{figure*}[tb!]
\begin{center}
\includegraphics[width=1\linewidth]{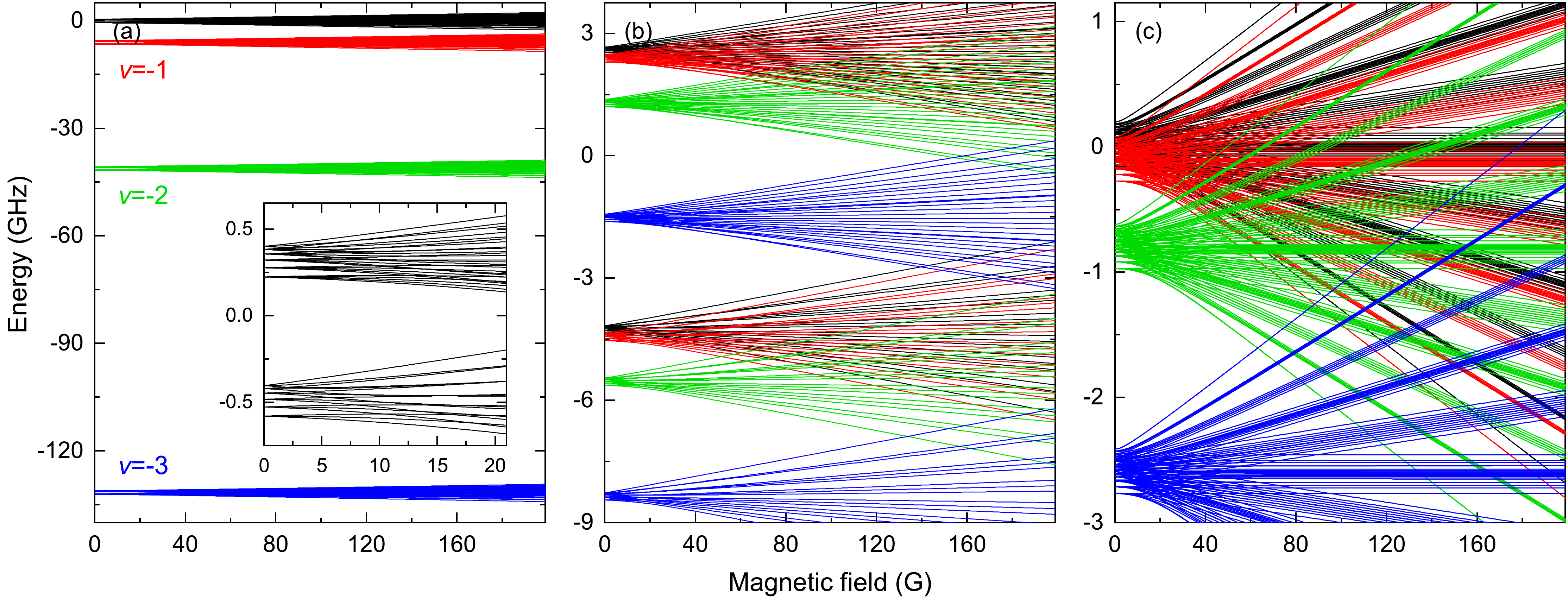}
\end{center}
\caption{Hyperfine energy levels for mixtures of (a)~\ce{^153Eu}+\ce{^7Li}, (b)~\ce{^153Eu}+\ce{^87Rb}, and (c)~\ce{^153Eu}+\ce{^151Eu} atoms with $M_\text{tot}=0$ as a function of the magnetic field. Black lines show atomic thresholds, whereas colorful (gray scale) lines correspond to the progression of the last three most weakly bound vibrational molecular levels for infinitely large and negative scattering lengths.}
\label{fig:hyperfine_EuX}
\end{figure*}

We perform \textit{ab initio} quantum scattering calculations using the coupled-channel formalism as implemented in Refs.~\cite{JanssenPRL13,TomzaPRL14,TomzaPRA15b}. We construct the total scattering wave function and Hamiltonian in a fully uncoupled basis set,
\begin{equation}\label{eq:basis}
|i_A,m_{i_A}\rangle
|s_A,m_{s_A}\rangle
|i_B,m_{i_B}\rangle
|s_B,m_{s_B}\rangle
|L,m_L\rangle\,,
\end{equation}
where $m_j$ is the projection of the angular momentum $j$ on the space-fixed $z$ axis, including all possible spin configurations, but assuming the projection of the total angular momentum $M_\mathrm{tot}=m_{f_A}+m_{f_B}+m_L=m_{i_A}+m_{s_A}+m_{i_B}+m_{s_B}+m_L$ to be conserved. Next, we transform the Hamiltonian to the basis of atomic hyperfine eigenstates, which is asymptotically diagonal. For homonuclear collisions of Eu atoms we impose properly the bosonic symmetry by transforming the wave function and Hamiltonian to the basis with well-defined total electronic spin $S$, nuclear spin $I$, and rotational $L$ angular momenta, and restricting the Hilbert space to its bosonic sector~\cite{HutsonPRA08,JanssenPRL13}. We solve the coupled-channel equations using a renormalized Numerov propagator~\cite{JohnsonJCP78} with step-size doubling and about 100 step points per de Broglie wavelength. The wave function ratio $\Psi_{i+1}/\Psi_{i}$ at the $i$--th grid step is propagated from small finite interatomic separations in the classically forbidden region where the scattering wave-function amplitude is negligible to large interatomic separations where electronic and dipolar potentials are negligible as compared to the collision energy. 
Then the $K$ and $S$ matrices are extracted by imposing the long-range scattering boundary conditions in terms of Bessel functions. The scattering lengths are obtained from the $S$ matrices for the lowest entrance channels $a_0=(1-S_{00})(1+S_{00})/(ik)$, where $k=\sqrt{2\mu E/\hbar^2}$ and $E$ is the collision energy. Feshbach resonances are characterized by their positions $B_0$, widths $\Delta$, and background scattering lengths $a_\text{bg}$, obtained by numerical fitting of the functional form  $a(B)=a_\text{bg}\left(1-\Delta/(B-B_0)\right)$ to the calculated scattering lengths in the vicinity of resonance poles. All calculations are carried out for the collision energy of 100$\,$nK.

Interaction between $^8S$-state Eu and $^2S$-state Li or Rb atoms results in two molecular electronic states of the $^7\Sigma^-$ and $^9\Sigma^-$ symmetries which have total electronic spin of $S=3$ and $4$, respectively. Interaction between two $^8S$-state Eu atoms results in eight  electronic states of the $^1\Sigma_g^+$, $^3\Sigma_u^+$, $^5\Sigma_g^+$, $^7\Sigma_u^+$, $^9\Sigma_g^+$, $^{11}\Sigma_u^+$, $^{13}\Sigma_g^+$, and $^{15}\Sigma_u^+$ symmetries with the total electronic spin $S$ from 0 to 7, respectively. The energy differences between molecular electronic states with different total electronic spin result from the exchange interaction.

For the Eu+Rb and Eu+Li systems we use potential-energy curves calculated in Ref.~\cite{TomzaPRA14}. Analytical potential-energy functions $V_{S}(R)$ are fitted to \textit{ab initio} data separately for $S=3$ and $4$ assuming the same long-range van der Waals coefficients $C_6$ reported in Ref.~\cite{TomzaPRA14}. For Eu+Eu system we use potential energy curves calculated in Ref.~\cite{BuchachenkoJCP09}. For this system, however, it was shown that the exchange interaction is small, and a family of potential-energy curves with different total electronic spin $S$ can be reproduced assuming the Heisenberg spin-exchange model of the spin-exchange interaction between $f$-shell electrons of Eu atoms~\cite{BuchachenkoJCP09}. As a result, the potential-energy curves for the Eu+Eu system read 
\begin{equation}
V_S(R)=V_{S=7}(R)+J(R)(56-S(S+1))/2\,,
\end{equation}
where functions $V_{S=7}(R)$ and $J(R)$ were calculated using \textit{ab initio} methods in Ref.~\cite{BuchachenkoJCP09}, and here we fit an analytical formula to them. 

Morse/Long-range potential-energy functions~\cite{LeRoyJCP06} are used to represent $V_{S}(R)$. They are given by
\begin{equation}
V_S(R)=D_e\left[1-\frac{u_\mathrm{LR}(R)}{u_\mathrm{LR}(R_e)}\exp\left(-\phi(R)y_p(R)\right)\right]^2-D_e\,,
\end{equation}
where $D_e$ and $R_e$ are the well depth and equilibrium distance of the interaction potential, respectively. The long-range part of the interaction potential is given by
\begin{equation}
u_\mathrm{LR}(R)=-\frac{C_6}{R^6}\,,
\end{equation}
whereas other functions are of the form
\begin{equation}
\begin{split}
y_p(R)&=\frac{R^p - R_e^p}{R^p + R_e^p}\,,\\
\phi(R)&= \varphi_\infty \, y_p(R) + \Big(1 - y_p(R)\Big) \sum_{i=0}^{4} \varphi_i y_q^i(R)\,,
\end{split}
\end{equation}
with $\varphi_\infty = \ln \left(\frac{2 D_e}{u_\mathrm{LR}(R_e)}\right)$, $p=4$, and $q=4$.
$D_e$, $R_e$, and $C_6$ are directly taken as reported in Refs.~\cite{TomzaPRA14,BuchachenkoJCP09}. The free parameters in the potential-energy functions, $\varphi_i$ ($i=0-4$), are determined by numerical fitting to the \textit{ab initio} points from Refs.~\cite{TomzaPRA14,BuchachenkoJCP09}. The obtained parameter values are presented in Table~\ref{tab:mlr}.
The $R$-dependent spin coupling constant $J(R)$ of the underlying Heisenberg model for the Eu+Eu system can be accurately approximated by the function
\begin{equation}
J(R)=\alpha/\cosh \left(\beta (R -R_0)\right)\,,
\end{equation}
where $\alpha=-0.53915\,$cm$^{-1}$, $\beta=0.79223\,$bohr$^{-1}$, and $R_0=7.8760\,$bohr~are obtained by numerical fitting to the \textit{ab initio} points from Ref.~\cite{BuchachenkoJCP09}. Such a function has a proper, exponentially decaying with $R$, asymptotic behavior.

We set the scattering lengths $a_S$ of the employed potential-energy curves by scaling them with appropriate factors $\lambda$, $V_S(R)\to \lambda V_S(R)$, taking values in the range of 0.97--1.03. We express the scattering lengths in the units of characteristic length scales of the van der Waals interaction, $R_6$, given by 
\begin{equation}
R_6=\left(\frac{2\mu C_6}{\hbar^2}\right)^{1/4}\,.
\end{equation}
It takes values 84$\,$bohr, 166$\,$bohr, and 178$\,$bohr for $^{153}$Eu+\ce{^7Li}, $^{153}$Eu+\ce{^87Rb}, and $^{153}$Eu+$^{151}$Eu, respectively. The corresponding characteristic energy scale is given by $E_6=\hbar^2/(2\mu R_6^2)$, and is 1.8$\,$mK, 56$\,\mu$K, and 36$\,\mu$K for the above mixtures.

\section{Numerical results and discussion}
\label{sec:results}

Before we discuss the results for magnetically tunable Feshbach resonances, we will analyze the hyperfine structures of the investigated mixtures and their impact on the scattering properties. Figure~\ref{fig:hyperfine_EuX} presents hyperfine energy levels for mixtures of $^{153}$Eu+$^7$Li, $^{153}$Eu+$^{87}$Rb, and $^{153}$Eu+$^{151}$Eu atoms with $M_\text{tot}=0$ as a function of the magnetic field. Black lines show atomic thresholds which are the result of combining the atomic hyperfine energy levels, presented in Fig.~\ref{fig:hyperfine}, summing to selected $M_\text{tot}$. The colorful (grayscale) lines are the atomic thresholds shifted by the largest possible binding energies of the last three most weakly bound vibrational levels supported by the van der Waals potentials determined by the long-range coefficients $C_6$~\cite{GaoPRA00}. These positions of molecular levels correspond to infinitely large and negative scattering lengths and true molecular binding energies must be equal to or smaller than them, and lie in such defined bins. The number of mixture's hyperfine energy levels is the largest for $M_\text{tot}=0$; therefore, the presented spectra correspond to the richest limiting cases, which are expected to be associated with the largest number of Feshbach resonances. Feshbach resonances can occur at crossings of molecular levels and atomic thresholds. 

Interestingly, the most important energy scale for the $^{153}$Eu+$^7$Li system is associated with the vibrational spacing, which is large because of the small reduced mass. At the same time, the hyperfine coupling constants are small for both atoms. In this case, the positions and properties of Feshbach resonances will crucially depend on the background scattering lengths and related binding energies of vibrational levels. For the $^{153}$Eu+$^{87}$Rb system, the energy scales associated with the vibrational and hyperfine structures are of similar order of magnitude, and the spectrum of Feshbach resonances will be a result of the interplay of both energy scales. For the $^{153}$Eu+$^{151}$Eu system, the vibrational binding energies are small because of the large reduced mass, and thus the properties of Feshbach resonances will depend crucially on the hyperfine structure, even though the hyperfine coupling constants are small in this system. As a result, it is guaranteed that a large number and density of resonances can be expected at relatively weak magnetic field strengths below 200$\,$G. The hyperfine spectrum for the homonuclear $^{153}$Eu+$^{153}$Eu system, which is equivalent to the heteronuclear one restricted to the bosonic sector of the Hilbert space, will have roughly twice smaller number of atomic and molecular levels, but other characteristics will be the same as in the heteronuclear case.

The positions and widths of Feshbach resonances depend on the hyperfine structure, progression of weakly bound rovibrational levels just below atomic thresholds, and background scattering lengths, as discussed above. Unfortunately, even the most accurate potential-energy functions obtained in the most advanced \textit{ab initio} electronic structure calculations do not allow one to predict accurately the scattering lengths for collisions between many-electron atoms, except for the systems with small number of bound states~\cite{KnoopPRA14}. Thus, at present, it is impossible to determine all parameters of Feshbach resonances without \textit{a priori} experimental knowledge. Nevertheless, the general characteristics of Feshbach resonances, such as the density of resonances and typical widths, can be learned by tuning the scattering lengths around the values of the characteristic length scales of the underlying van der Waals interactions $R_6$. Therefore, we have calculated the spectra of magnetic Feshbach resonances for a large number of combinations of scattering lengths and present the most typical ones.
 
\begin{figure*}[p]
\centering
\includegraphics[width=\linewidth]{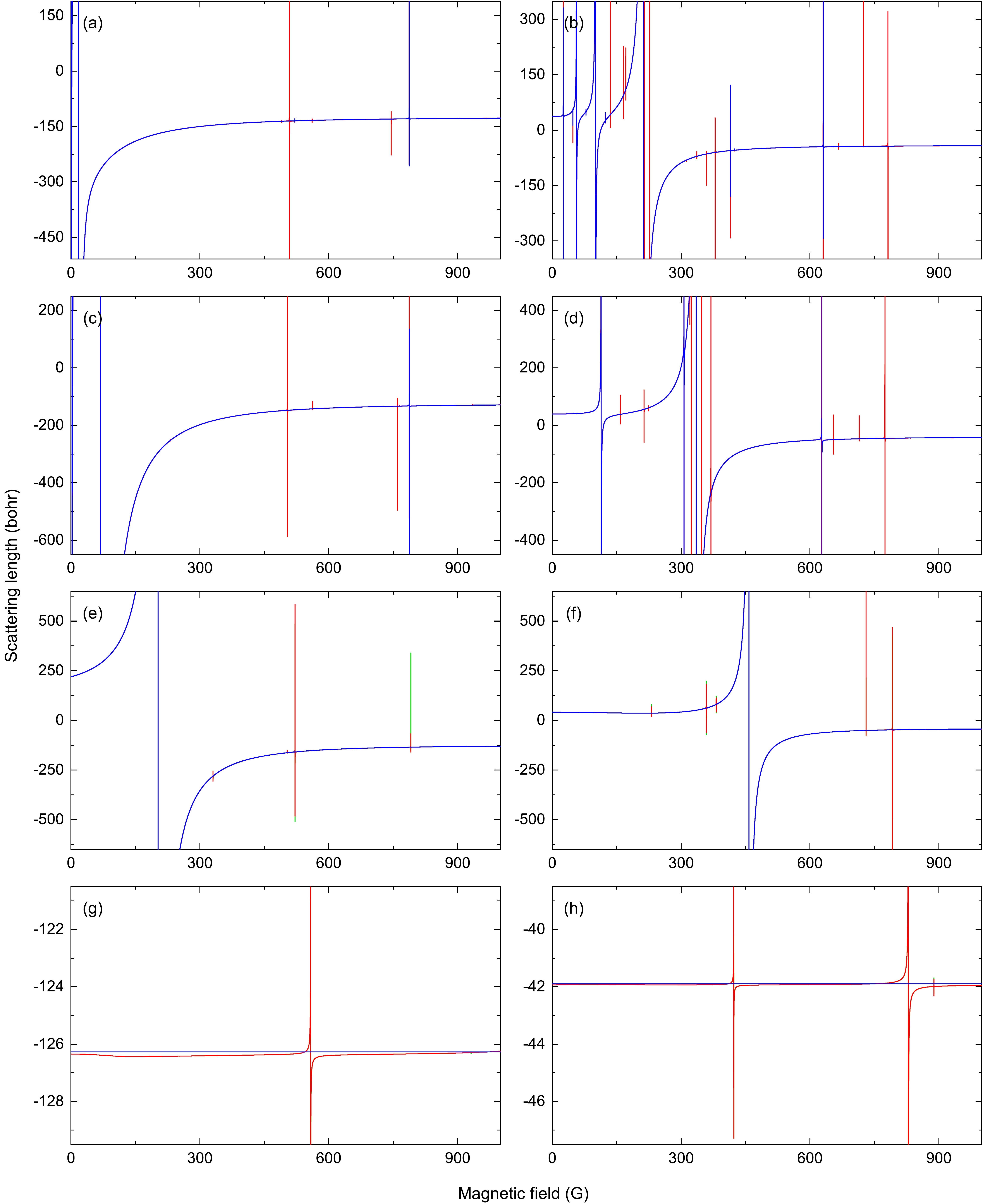}
\caption{Scattering lengths for ultracold collisions between \ce{^153Eu} and \ce{^7Li} atoms as a function of the magnetic field: (a),(b) for $M_\mathrm{tot}=0$, (c),(d) for $M_\mathrm{tot}=-6$, (e),(f) for $M_\mathrm{tot}=-7$, and (g),(h) for $M_\mathrm{tot}=-8$. The following scattering lengths for the potential-energy functions are assumed: (a),(c),(e),(g) $a_{S=3}=1.5 R_6$ and  $a_{S=4}=-1.5 R_6$, (b),(d),(f),(h) $a_{S=3}=0.5 R_6$ and  $a_{S=4}=-0.5 R_6$. Blue (dark gray) lines show scattering lengths without the dipole-dipole interaction included and red (gray) and green (light gray) lines show scattering lengths with the dipole-dipole interaction included with $L_\text{max}=2$ and $4$, respectively. Note different scales for different $M_\mathrm{tot}$.}
\label{fig:EuLi}
\end{figure*}

\begin{figure*}[p]
\centering
\includegraphics[width=1\linewidth]{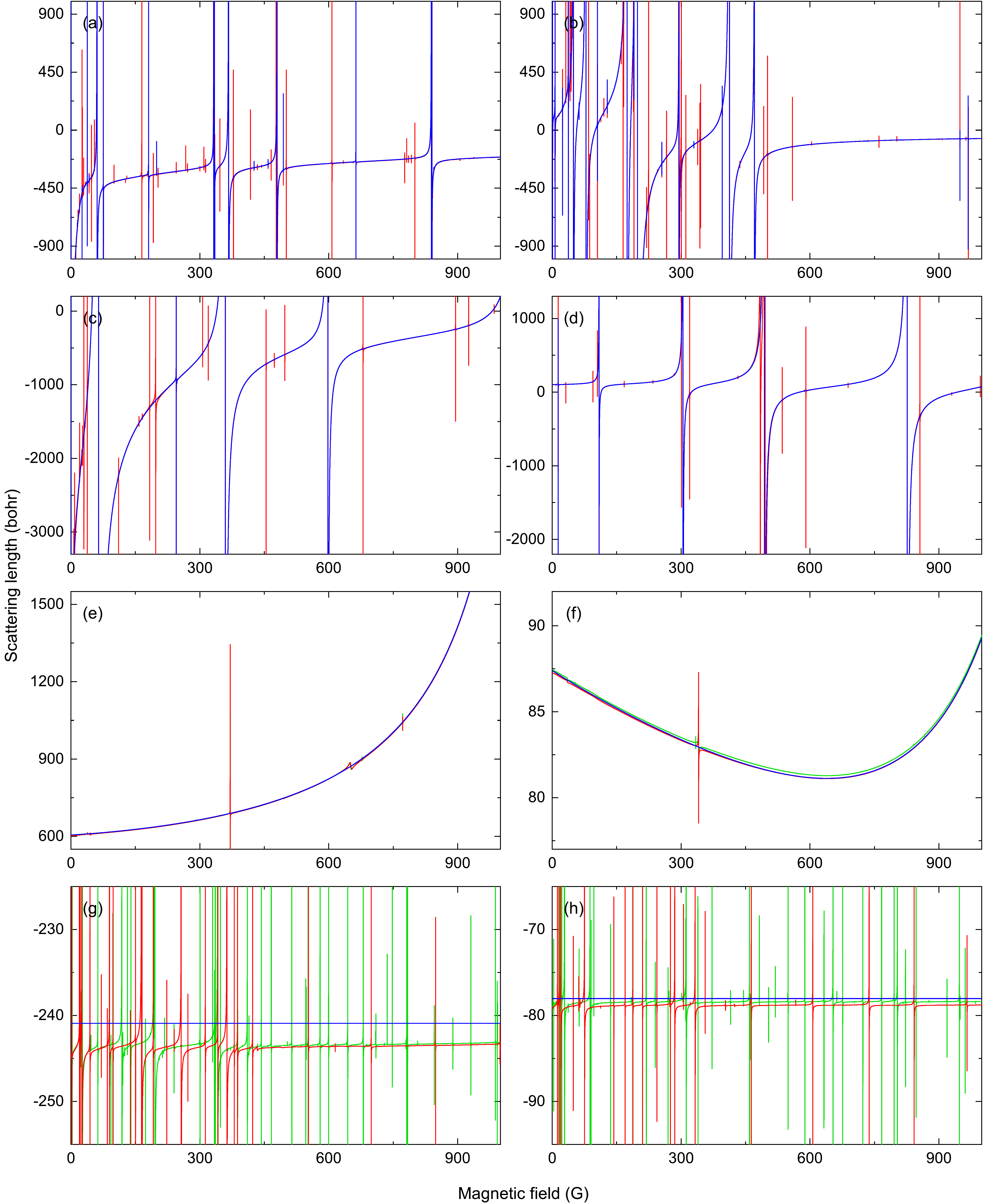}
\caption{Scattering lengths for ultracold collisions between \ce{^153Eu} and \ce{^87Rb} atoms as a function of the magnetic field: (a),(b) for $M_\mathrm{tot}=0$, (c),(d) for $M_\mathrm{tot}=-6$, (e),(f) for $M_\mathrm{tot}=-7$, and (g),(h) for $M_\mathrm{tot}=-8$. The following scattering lengths for the potential-energy functions are assumed: (a),(c),(e),(g) $a_{S=3}=1.5 R_6$ and  $a_{S=4}=-1.5 R_6$, (b),(d),(f),(h) $a_{S=3}=0.5 R_6$ and  $a_{S=4}=-0.5 R_6$. Blue (dark gray) lines show scattering lengths without the dipole-dipole interaction included and red (gray) and green (light gray) lines show scattering lengths with the dipole-dipole interaction included with $L_\text{max}=2$ and $4$, respectively. Note different scales for different $M_\mathrm{tot}$.}
\label{fig:EuRb}
\end{figure*}

\begin{figure*}[p]
\centering
\includegraphics[width=1\linewidth]{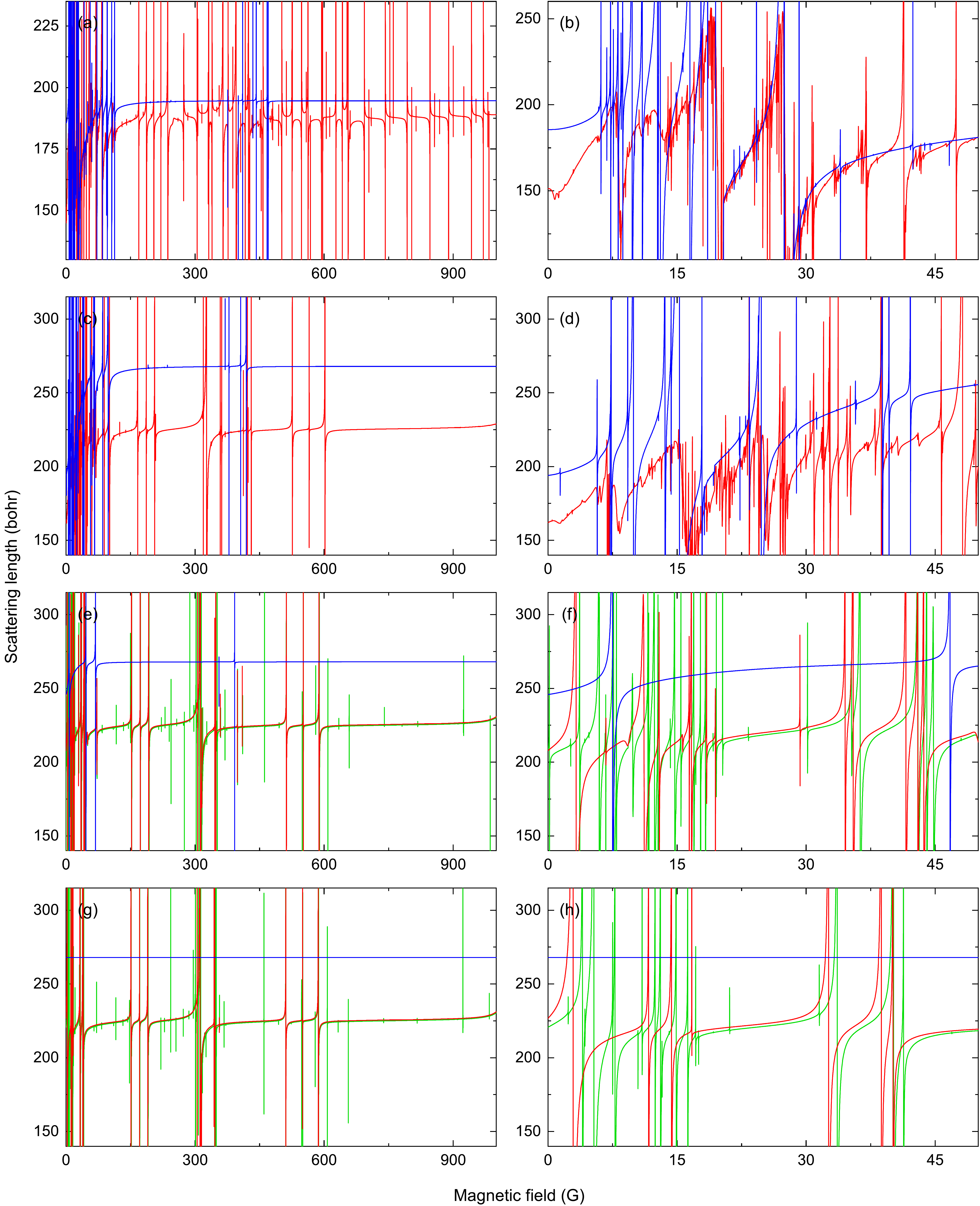}
\caption{Scattering lengths for ultracold collisions between \ce{^153Eu} and \ce{^151Eu} atoms as a function of the magnetic field: (a),(b) for $M_\mathrm{tot}=0$, (c),(d) for $M_\mathrm{tot}=-5$, (e),(f) for $M_\mathrm{tot}=-11$, and (g),(h) for $M_\mathrm{tot}=-12$. Panels (b),(d),(f),(g) are zoomed versions of panels (a),(c),(e),(h). The scattering length of $a_{S=7}=1.5 R_6$ is assumed for spin-polarized collisions. Blue (dark gray) lines show scattering lengths without the dipole-dipole interaction included and red (gray) and green (light gray) lines show scattering lengths with the dipole-dipole interaction included with $L_\text{max}=2$ and $4$, respectively.}
\label{fig:EuEu151}
\end{figure*}

\begin{figure*}[p]
\centering
\includegraphics[width=1\linewidth]{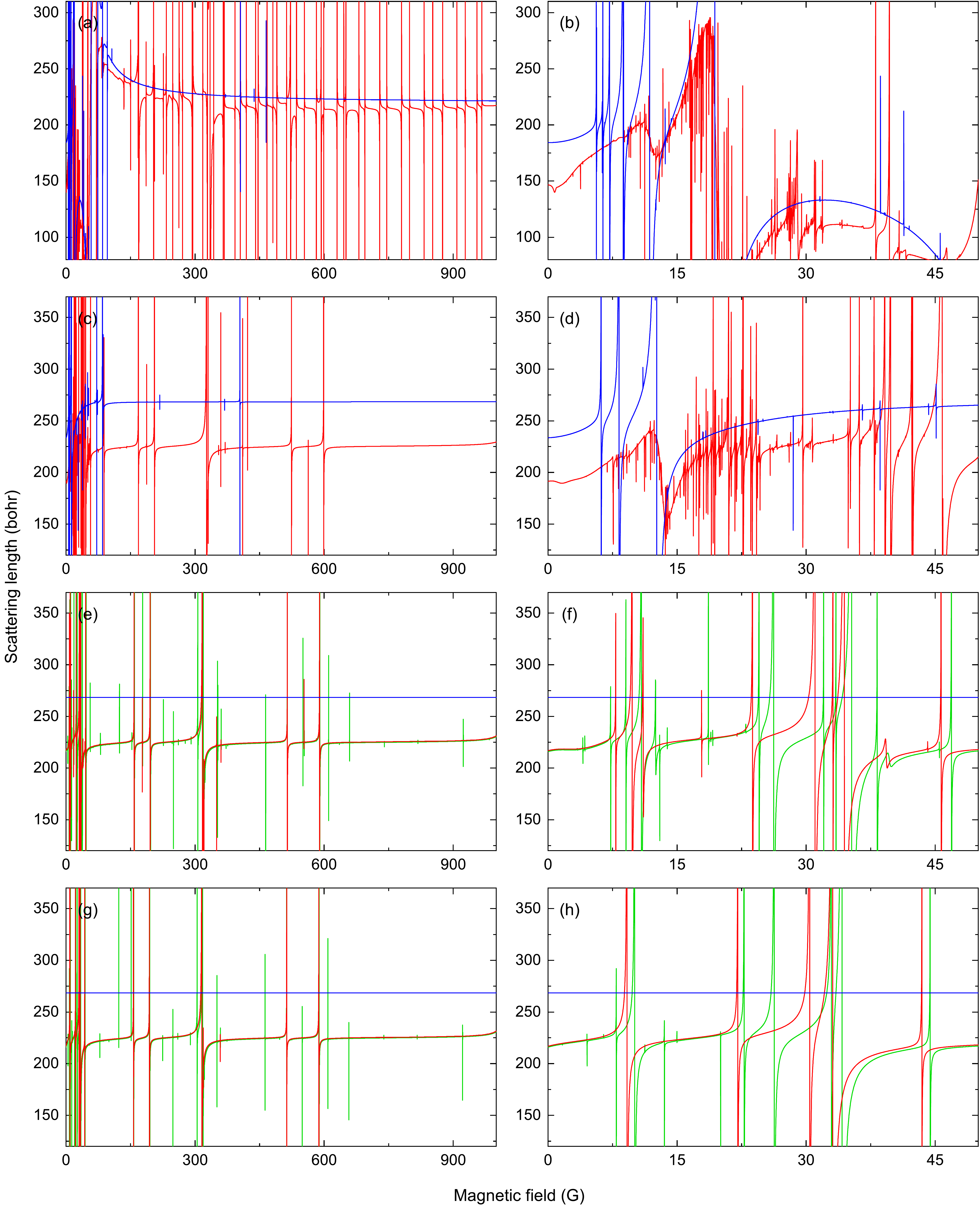}
\caption{Scattering lengths for ultracold collisions between \ce{^153Eu} atoms as a function of the magnetic field: (a),(b) for $M_\mathrm{tot}=0$, (c),(d) for $M_\mathrm{tot}=-5$, (e),(f) for $M_\mathrm{tot}=-11$, and (g),(h) for $M_\mathrm{tot}=-12$. Panels (b),(d),(f),(g) are zoomed versions of panels (a),(c),(e),(h). The scattering length of $a_{S=7}=1.5 R_6$ is assumed for spin-polarized collisions. Blue (dark gray) lines show scattering lengths without the dipole-dipole interaction included and red (gray) and green (light gray) lines show scattering lengths with the dipole-dipole interaction included with $L_\text{max}=2$ and $4$, respectively.}
\label{fig:EuEu153}
\end{figure*}

Figures~\ref{fig:EuLi}--\ref{fig:EuEu153} show $s$-wave scattering lengths for ultracold collisions in the $^{153}$Eu+$^7$Li,  $^{153}$Eu+$^{87}$Rb, $^{153}$Eu+$^{151}$Eu, and $^{153}$Eu+$^{153}$Eu systems as a function of the magnetic-field strength. Results are presented for several different projections of the total angular momentum on the magnetic field $M_\mathrm{tot}$, including collisions with $M_\mathrm{tot}=0$, which correspond to the largest number of channels, and ones with maximal possible $|M_\mathrm{tot}|$ that correspond to maximally spin-stretched states, for which only resonances induced by the dipole-dipole interaction can occur. In all calculations the same step in the magnetic-field strength of 0.01$\,$G is assumed; therefore, the prominence of the resonances can be related to their widths, which can be visually compared between different channels and systems. Different colors encode results without the dipole-dipole interaction included, that are obtained by restricting the basis set given by Eq.~\eqref{eq:basis} to $L_\text{max}=0$, and with the dipole-dipole interaction included when $L_\text{max}=2$ or $4$. For $L_\text{max}=2$ $d$-wave resonances appear, and for $L_\text{max}=4$ additionally $g$-wave resonances emerge. These resonances are a result of coupling of the $s$-wave entrance channel with $d$-wave and $g$-wave bound molecular levels in closed channels. The coupling with $d$-wave bound levels is direct, whereas the coupling with $g$-wave bound levels is indirect via $d$-wave channels. Calculations including $L_\text{max}=4$ are presented only for channels with large $|M_\mathrm{tot}|$ for the clarity and because the $g$-wave resonances are at least an order of magnitude narrower than the $d$-wave ones. At the same time, the number of channels increases fast with $L_\text{max}$. For example, for $M_\mathrm{tot}=0$ of the $^{153}$Eu+$^{7}$Li or $^{153}$Eu+$^{87}$Rb system the number of channels is 46, 258, and 582 for $L_\text{max}=0$, $2$, and $4$, respectively, whereas for $M_\mathrm{tot}=0$ of the $^{153}$Eu+$^{151}$Eu system the number of channels is 218, 1252, and 2916 for $L_\text{max}=0$, $2$, and $4$, respectively.

The dipole-dipole interaction not only couples different partial waves but it also modifies the long-range character of the interatomic interaction potential in higher partial waves from $1/R^6$ to $1/R^3$~\cite{MarinescuPRL98,MelezhikPRL03,BaranovPR08}. Thus, in the zero collision energy limit, not only $s$-wave scattering length but also $d$-wave one can have a finite value. We present only $s$-wave scattering lengths for brevity and because they are at least an order of magnitude larger than $d$-wave ones in the present case.


Figures~\ref{fig:EuLi} and~\ref{fig:EuRb} show $s$-wave scattering lengths for ultracold collisions in the $^{153}$Eu+$^7$Li and  $^{153}$Eu+$^{87}$Rb systems as a function of the magnetic-field strength. Results are presented for two sets of scattering lengths: $a_{S=3}=1.5 R_6$ and  $a_{S=4}=-1.5 R_6$, and $a_{S=3}=0.5 R_6$ and  $a_{S=4}=-0.5 R_6$. The first set corresponds to rather large and favorable for broad resonances scattering lengths, whereas the second one corresponds to rather small and less favorable scattering lengths. As expected, the largest number of resonances is observed for collisions with $M_\mathrm{tot}=0$, counting around 10 and 30 $s$-wave resonances, and 30 and 100 $d$-wave resonances below 1000$\,$G for the $^{153}$Eu+$^7$Li and  $^{153}$Eu+$^{87}$Rb mixtures, respectively. The number of resonances decreases with increasing $|M_\mathrm{tot}|$ and there are no $s$-wave resonances for fully spin-polarized collisions with $M_\mathrm{tot}=-8$ for which, however, higher wave resonances exist. Interestingly, for the spin-polarized $^{153}$Eu+$^{87}$Rb mixture, there are around 30 $d$-wave resonances and around 200 $g$-wave resonances below 1000$\,$G. The broad $s$-wave Feshbach resonances have widths around 10-100$\,$G for $^{153}$Eu+$^7$Li and around 1-10$\,$G for $^{153}$Eu+$^{87}$Rb. The $d$-wave Feshbach resonances have widths around 10-100$\,$mG and $g$-wave Feshbach resonances have widths below 10$\,$mG for both systems. The $s$-wave resonances in mixtures of europium and alkali-metal atoms have a very similar nature to the resonances between alkali-metal atoms because the exchange-interaction-induced splitting between two electronic states is relatively large. At the same time, higher wave resonances are expected to be broader in the present case because the dipole-dipole interaction is seven times stronger between europium and alkali-metal atoms than between alkali-metal atoms.

\begin{figure}[tb!]
\begin{center}
\includegraphics[width=1\linewidth]{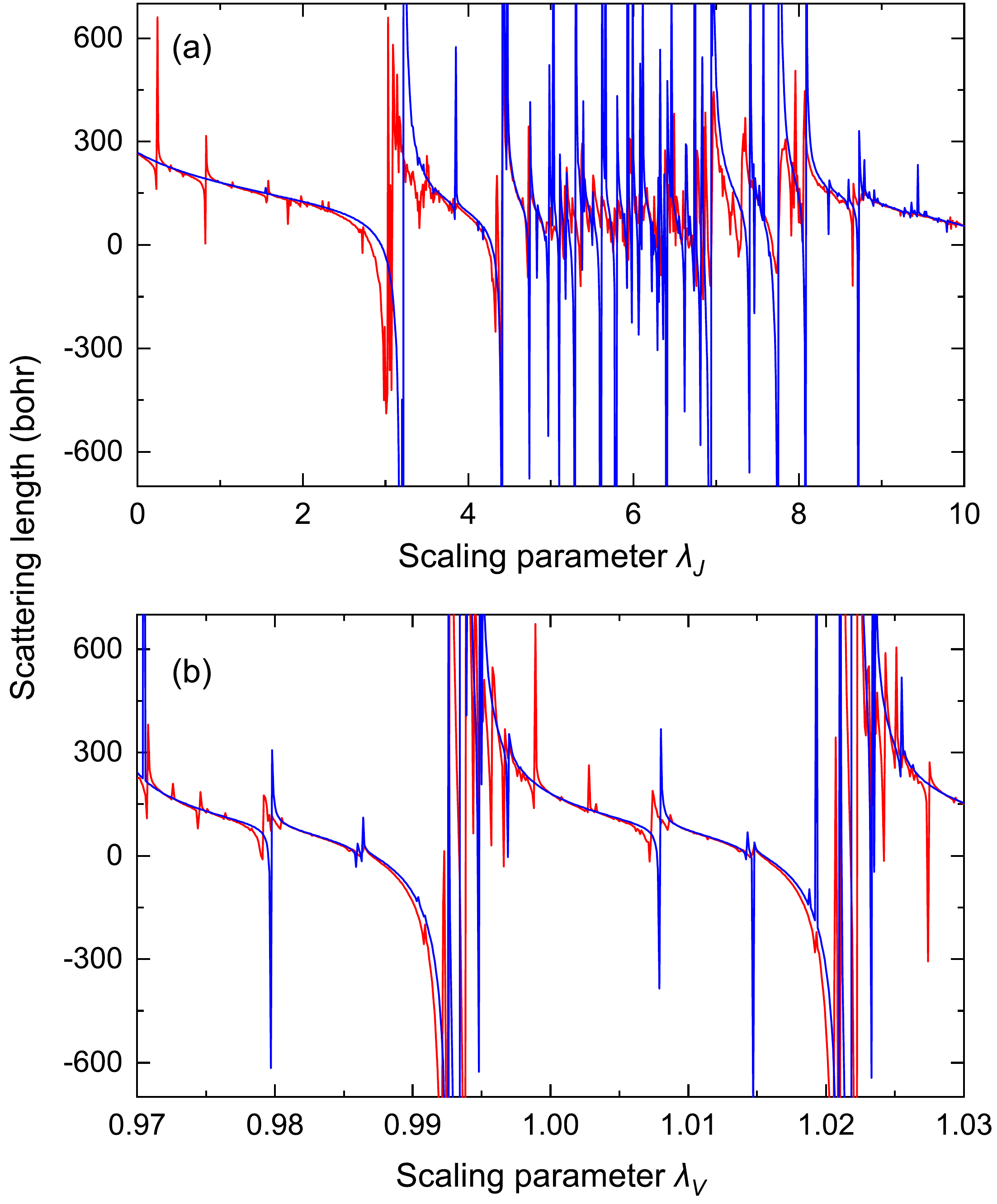}
\end{center}
\caption{Scattering lengths for ultracold collisions between \ce{^153Eu} and \ce{^151Eu} atoms with $M_\text{tot}=0$ at the magnetic field $B=50\,$G as a function of the parameters scaling the anisotropic (a) $J(R)\rightarrow\lambda_J J(R)$ and isotropic (b) $V_{S=7}(R)\rightarrow\lambda_V V_{S=7}(R)$ interatomic electronic interaction potential functions. The scattering length of $a_{S=7}=1.5 R_6$ is assumed. Blue (dark gray) lines show scattering lengths without the dipole-dipole interaction included and red (gray) lines show scattering lengths with the dipole-dipole interaction included.}
\label{fig:scaling}
\end{figure}

Figures~\ref{fig:EuEu151} and~\ref{fig:EuEu153} show $s$-wave scattering lengths for ultracold collisions in the $^{153}$Eu+$^{151}$Eu and $^{153}$Eu+$^{153}$Eu systems as a function of the magnetic field strength. The scattering length for the electronic potential energy curve with the total electronic spin of $S=7$, which governs the spin-polarized collisions, is set to $a_{S=7}=1.5 R_6$. There are around 100 $s$-wave and 200 $d$-wave Feshbach resonances below 1000$\,$G for the collisions with $M_\mathrm{tot}=0$, and this number slowly decreases with increasing $|M_\mathrm{tot}|$. At the same time, two-thirds of resonances are located below 200$\,$G because of the small hyperfine coupling constants for Eu atoms [cf.~\ref{fig:hyperfine_EuX}(c)]. For small $|M_\text{tot}|$, the spectra below 200$\,$G are very dense with many overlapping resonances and the density of resonances approaches one per Gauss. The number of resonances for the homonuclear combination is smaller than for the heteronuclear mixture, but the reduction in the number of visible resonances is smaller than the reduction in the number of channels. The typical widths of both $s$-wave and $d$-wave resonances are between 10$\,$mG and 100$\,$mG, whereas the widths of $g$-wave resonances are below 10$\,$mG for both homonuclear and heteronuclear collisions. The inclusion of $g$-wave channels noticeably moves positions of $d$-wave resonances at small magnetic field strengths. Additionally, the dipole-dipole interaction visibly modifies the background scattering length (by up to around 20\%). Interestingly, the widths of $s$-wave resonances induced by the relatively weak short-range spin-exchange interaction and the widths of $d$-wave resonances induced by the relatively strong long-range magnetic dipole-dipole interaction are of the same order of magnitude. In fact, the spin-exchange interaction between Eu atoms counts below $0.1\,\%$ of the total electronic interaction energy at the equilibrium geometry and was classified as extremely weak as compared to typical energy scales of the exchange interaction in other molecular systems~\cite{BuchachenkoJCP09}.

Due to its very small value and computational complexity, the spin-exchange interaction calculated for the Eu+Eu system~\cite{BuchachenkoJCP09} is the most uncertain parameter of our model. It was already shown that its actual value is crucial to determine correctly the Zeeman relaxation rates for collisions of magnetically trapped Eu atoms~\cite{SuleimanovPRA10}. Therefore, we have also evaluated ultracold collisions between Eu atoms as a function of the magnetic-field strength for several values of the scaling parameter $\lambda_J$, where linear scaling of the spin-exchange interaction $J(R)\rightarrow\lambda_J J(R)$ was assumed. An exemplary dependence of the scattering lengths for ultracold collisions between $^{153}$Eu and $^{151}$Eu atoms with $M_\text{tot}=0$ at the magnetic field strength of $B=50\,$G on the scaling parameter $\lambda_J$ is presented in Fig.~\ref{fig:scaling}(a). For $\lambda_J=1$ the dependence is weak and linear, only interrupted by $d$-wave resonances. That suggests the perturbative impact of the spin-exchange interatomic interaction on the collisions and $s$-wave resonances. The non-perturbative regime can be identified for the spin-exchange interaction increased three times or more. For $\lambda_J>3$ the number and density of resonances increase by a factor of two and stop to depend on $\lambda_J$. For example, for $\lambda_J=5$ the number of $s$-wave and $d$-wave resonances below 1000$\,$G for ultracold collisions between $^{153}$Eu and $^{151}$Eu atoms with $M_\text{tot}=0$ is 300 and 300, respectively. For comparison, Fig.~\ref{fig:scaling}(b) presents the dependence of the scattering lengths on the scaling of the isotropic part of the interaction potential which, as expected, is very strong.

We have observed similar characteristics as those presented in Figs.~\ref{fig:EuLi}-\ref{fig:scaling} also for different sets of scattering lengths. Only accidentally very close values of the background scattering lengths for the potential-energy curves with $S=3$ and $4$ can significantly reduce the widths of $s$-wave Feshbach resonances in mixtures of europium and alkali-metal atoms, while the widths of higher wave resonances can be reduced in all atomic combinations only if the scattering lengths for all the potential-energy curves are very close to zero. This is very improbable and can be resolved by changing the used isotopes.
Thus, for all the investigated systems, for a broad range of possible scattering lengths, there should exist, at least for some $M_\text{tot}$, favorable resonances for controlling ultracold collisions and magnetoassociation at magnetic-field strengths below 1000$\,$G. For mixtures of europium and alkali-metal atoms $s$-wave resonances as broad as between alkali-metal atoms (with widths much over 1$\,$G) can be expected. For ultracold homo- and heteronuclear gases of europium atoms a large number of useful $s$-wave and $d$-wave resonances (with widths reaching 100$\,$mG) can be expected even at magnetic field strengths below 100$\,$G. At the same time, it should be possible to find magnetic field strengths at which independent control of scattering properties in different scattering channels can be realized without being disturbed by accidental resonances. This should be a favorable condition for investigating magnetic polaron and similar phenomena in ultracold highly magnetic gases.  

\begin{figure}[tb!]
\begin{center}
\includegraphics[width=1\linewidth]{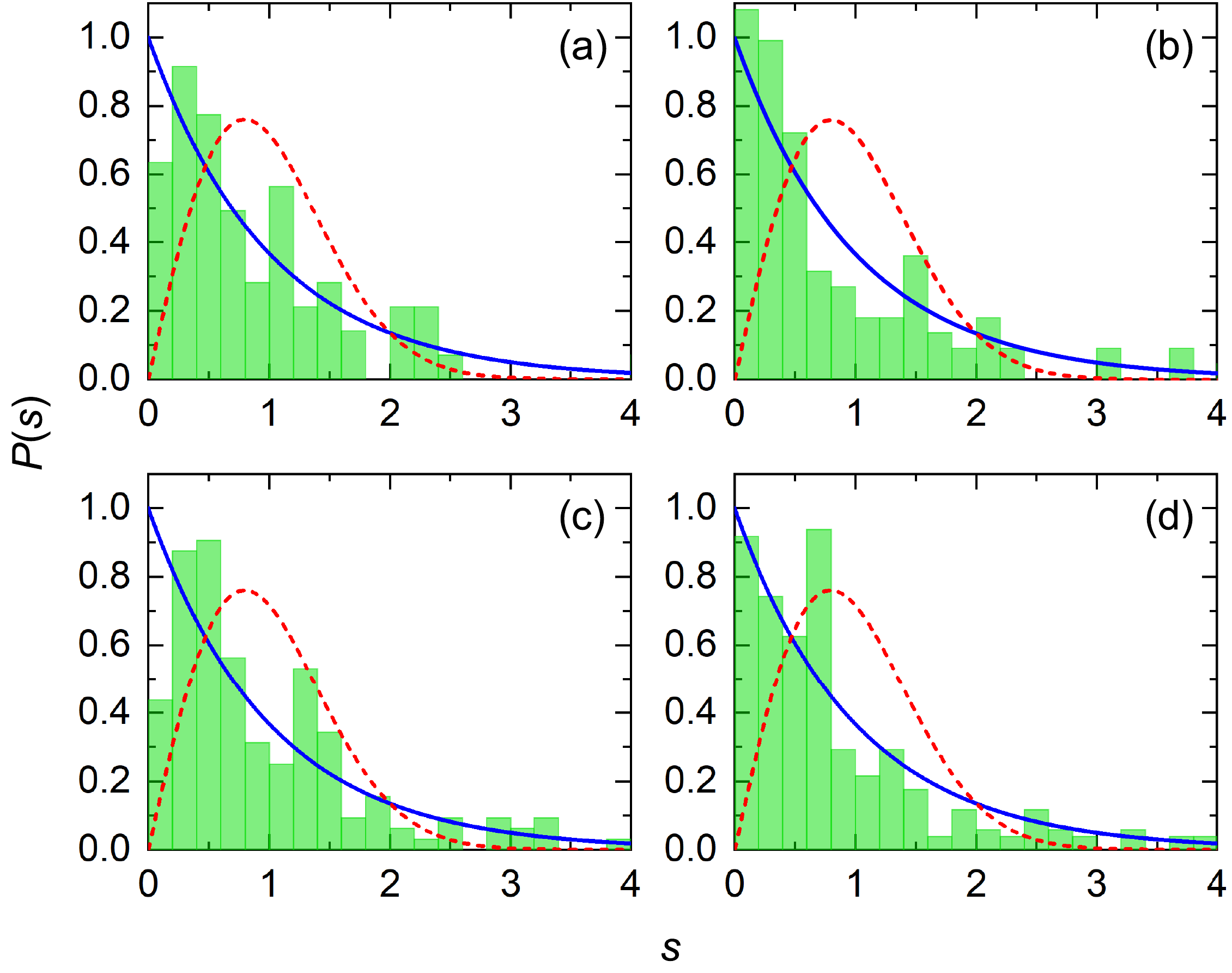}
\end{center}
\caption{Nearest-neighbor spacing distributions of $s$-wave (a),(c) and both $s$-wave and $d$-wave (b),(d) resonance positions for ultracold collisions between $^{153}$Eu and $^{151}$Eu atoms with $M_\text{tot}=0$ at magnetic field strengths between 0 and 100$\,$G with the spin-exchange interaction as obtained in \textit{ab initio} calculations (a),(b) and scaled to the non-perturbative regime by $\lambda_J=5$ (c),(d). The Poisson (solid blue) and Wigner-Dyson (dashed red) distribution curves are plotted for comparison.}
\label{fig:histograms}
\end{figure}

The complex spectra of many overlapping Feshbach resonances observed in Figs.~\ref{fig:EuEu151}(a) and ~\ref{fig:EuEu151}(b) and Figs.~\ref{fig:EuEu153}(a) and ~\ref{fig:EuEu153}(b) raise a question of whether the investigated systems exhibit a quantum chaotic behavior. For ultracold collisions of Dy and Er atoms it was measured~\cite{FrischNature14,MaierPRX15} and theoretically confirmed~\cite{MaierPRX15,JachymskiPRA15,GonzalezPRA15,GreenPRA16,YangPRL17,MakridesSA18} that the interplay of anisotropic electronic and dipolar interactions leads to the chaotic spectra of Feshbach resonances being the signature of the level repulsion following the predictions of the Gaussian Orthogonal Ensemble of random matrices~\cite{Haake2010}. Similar results were predicted for atom-molecule collisions~\cite{FryePRA16,CroftNC17,CroftPRA17}. To verify the above hypothesis in the considered case, in Fig.~\ref{fig:histograms} we present nearest-neighbor spacing distributions of Feshbach resonance positions for ultracold collisions between $^{153}$Eu and $^{151}$Eu atoms with $M_\text{tot}=0$ at the magnetic field strengths between 0 and 100$\,$G, for which the density of overlapping resonances is the largest and chaotic behavior is the most probable. Although we have selected the most dense part of the spectrum, the number of resonances is still relatively small, which makes our analysis semi--quantitative. The distribution of uncorrelated energy levels should be described by the Poisson distribution, $P_\text{P}(s)=\exp(-s)$, whereas the quantum chaotic distribution should be described by the Wigner-Dyson distribution, $P_\text{WD}(s)=\frac{\pi s}{2}\exp({-\pi s^2/4})$, where the distance between adjacent levels $s$ is in the units of mean resonance spacing~\cite{Haake2010}. The transition between Poissonian and quantum chaotic Wigner-Dyson distributions can be quantified by the intermediate Brody distribution, $P_\text{B}(s,\eta)=b(1+\eta)s^\eta\exp(-bs^{\eta+1})$, with associated Brody parameter $\eta$~\cite{BrodyLNC73}, which is 0 for Poisson and 1 for Wigner-Dyson distribution. In the present case, for the distribution of $s$-wave resonance positions, the level repulsion can be noticed, but the Brody parameter does not exceed 0.25. When $d$-wave resonances are included in the spectra, the level repulsion is less pronounced with the Brody parameter not exceeding 0.1. Very similar results are obtained for both native and increased spin-exchange interaction, with only very slight increase of the Brody parameter in the second case. This suggests that the Heisenberg model describing the interatomic spin-exchange interaction in the Eu+Eu system does not support quantum chaotic behavior and the anisotropic interaction related to the non--zero electronic orbital angular momentum, as in Dy and Er, is needed~\cite{MaierPRX15}. The decrease of the level repulsion when the resonances induced by the dipole-dipole interaction are included agrees with previous theoretical works~\cite{MaierPRX15,YangPRL17}, which show that the magnetic dipole moment of Dy and Er, and so of Eu, is too small to support quantum chaotic behavior on its own. For larger $|M_\text{tot}|$ and for mixtures of europium with alkali-metal atoms the resonance spectra are too simple to expect quantum chaotic signatures.

\section{Summary and conclusions}
\label{sec:summary}

Motivated by recent advances in production and application of ultracold highly dipolar atoms in complex electronic states, such as Er and Dy~\cite{LevPRL11,LevPRL12,FerlainoPRL12,AikawaPRL14,PetrovPRL12,MaierPRX15,MaierPRA15}, we have considered ultracold collisions involving Eu atoms as another lanthanide candidate for the realization and application of dipolar atomic and molecular quantum gases in many-body physics. Dy and Er atoms are excellent systems for experiments exploiting dipolar interactions, but their very complex internal structure resulting in very rich, dense, and chaotic spectra of unavoidable magnetic Feshbach resonances~\cite{MaierPRA15} can limit applications based on the precision control of internal degrees of freedom, such as magnetoassociation, optical stabilization to deeply bound states, or magnetic polaron physics investigations.

Therefore, we have investigated magnetically tunable Feshbach resonances between ultracold europium atoms and between europium and alkali-metal atoms using multichannel quantum scattering calculations. We have studied both homonuclear $^{153}$Eu+$^{153}$Eu and heteronuclear $^{151}$Eu+$^{153}$Eu systems of europium atoms and $^{153}$Eu+$^{87}$Rb and $^{153}$Eu+$^7$Li combinations. We have analyzed the prospects for the control of scattering properties, observation of quantum chaotic behavior, and magnetoassociation into ultracold polar and paramagnetic molecules. 

The most important of our findings can be summarized as follows.

(1) Favorable resonances are expected at experimentally feasible magnetic-field strengths below 1000$\,$G for all investigated atomic combinations.

(2) The density of resonances depends strongly on the projection of the total angular momentum on the magnetic field (the degree of polarization). 

(3) The dipole-dipole interaction between europium and alkali-metal atoms is weaker than the spin-exchange interaction; therefore, $s$-wave resonances are more  favorable than $d$-wave ones in these systems.    

(4) The dipole-dipole interaction between europium atoms is comparable to relatively weak short-range spin-exchange interaction, but strong enough to induce favorable resonances.  

(5) Large number and density of $s$-wave and $d$-wave resonances is expected in ultracold gases of europium atoms.

(6) Especially large number and density of resonances is expected at magnetic-field strengths below 200$\,$G, but signatures of quantum chaotic behavior measured by level repulsion are limited. 
 
The present results draw attention to Eu atoms as an interesting and favorable candidate for dipolar many-body physics and pave the way towards experimental studies and application at ultralow temperatures.

\begin{acknowledgments}
We would like to thank Mariusz Semczuk and Anna Dawid for useful discussions, and Alexei Buchachenko for providing us numerical values of the potential energy curves for the Eu$_2$ dimer. We acknowledge financial support from the Foundation for Polish Science within the Homing programme co-financed by the European Regional Development Fund, the National Science Centre Poland (2016/23/B/ST4/03231 and 2017/25/B/ST4/01486), and the PL-Grid Infrastructure. The work reported here was initiated during The Toru\'n Physics Summer Program 2017.
\end{acknowledgments}

\bibliography{EuX}

\end{document}